\documentstyle{mn}           		
\def\approxgt{\mathrel{\hbox{\rlap{\lower.55ex \hbox {$\sim$}}\kern-.3em \raise.4ex \hbox{$>$}}}}
\def \src {XB\thinspace1916$-$053}

\input epsf.sty
\begin{document}

\title[Progressive covering in XB\thinspace 1916-053]
{Progressive Covering of the Accretion Disk Corona 
during Dipping in the LMXB XB\thinspace 1916-053}
   \author[R. Morley et al.]
{R. Morley$^1$, M. J. Church$^1$, A. P. Smale$^2$
    and M. Ba\l uci\'nska-Church$^1$\\
   $^1$School of Physics and Astronomy, University of Birmingham,
   Birmingham, B15 2TT, UK\\
e-mail: rm@star.sr.bham.ac.uk, mjc@star.sr.bham.ac.uk, mbc@star.sr.bham.ac.uk\\
   $^2$Laboratory for High Energy Astrophysics, Code 660.2,
   NASA/Goddard Space Flight Center, MD 20771, USA\\
e-mail: alan@osiris.gsfc.nasa.gov}
\date{Accepted. Received}

   \maketitle

\begin{abstract}
Results are reported for analysis of the extensive {\it Rosat}
observation of the dipping low mass X-ray binary XB\thinspace 1916-053.
Dipping is 100\% deep showing that the emission regions are completely covered by
the absorber. A good fit to the non-dip spectrum is obtained using a model 
consisting of a blackbody with $\rm {kT_{BB}}$ = 1.95$\rm
{^{+0.74}_{-0.34}}$~keV and a power law with photon index
2.32$\pm0.04$. These components are identified with emission from the
neutron star, and Comptonized emission from an extended accretion disk 
corona (ADC). Dip spectra are well-fitted by rapid absorption of the blackbody,
and progressive covering of the extended component, as the absorber moves 
across the source, with a covering fraction that increases smoothly from 
zero to $\sim$1.0. Progressive covering shows that the Comptonized
emission region is extended, consistent with it originating in the accretion 
disk corona. The strong unabsorbed component in the dip spectra is 
well-modelled as the uncovered part of the Comptonized emission at all 
stages of dipping. There is no detectable change in the low 
energy cut-off of the spectrum in dipping which supports the identification
of the unabsorbed part of the spectrum with the uncovered part of the 
ADC emission. The absorbed part of the ADC emission
is rapidly removed from the 0.1--2.0~keV band of the {\it PSPC},
which therefore selects only the uncovered part of the emission, and 
so the spectral evolution in dipping as viewed by the {\it PSPC}
depends only on the covering fraction, determined by the geometric 
overlap between the source and absorber. 
\end{abstract}


\section{Introduction}
XB\thinspace 1916-053 is a member of the class of Low Mass X-ray Binary 
(LMXB) sources which exhibit dips in X-ray intensity at the orbital
period. It is generally
accepted that dipping is caused by absorption in the bulge in the outer 
accretion disk where the accretion flow from the companion star impacts
(White \& Swank 1982). \src\ has the shortest period of all dipping sources
of 50 min, and is also remarkable in having a difference of $\sim$1\% between
the X-ray and optical periods (Grindlay et al. 1988). Dipping is often 
very deep in this source, with the depth of dipping reaching $\sim $100\% 
in the band 1--10~keV.
The source has been observed using several X-ray observatories,
notably {\it Exosat}, {\it Ginga}, {\it ASCA} and {\it Rosat}. Spectral 
analysis of the three \hbox {{\it Exosat}} observations (Smale et al. 1988) showed that in dipping, part of the emission was clearly not absorbed.
The non-dip emission was well fitted by a simple power law
model. However, the dip spectra clearly contained two components: one absorbed
and the other not absorbed. These were modelled by dividing the non-dip model
into two parts each having the same form as the non-dip spectrum, one 
absorbed, and the other not absorbed, but with a normalization that decreased
strongly in dipping. This approach may be called the ``absorbed plus
unabsorbed'' approach, and has also been used in fitting the spectra 
of the dip sources XB\thinspace 1254-690
(Courvoisier et al. 1986) and XBT\thinspace 0748-676 (Parmar et al. 1986).
More recently {\it Ginga} data on \src\ has also been
fitted by this approach (Smale et al. 1992; Yoshida et al. 1995).
The parameters of the emission regions cannot, of course, change during 
dip intervals, and so the decrease in normalization of the unabsorbed 
component cannot represent a decrease in brightness of the source, and 
a difficulty in using ``absorbed plus unabsorbed'' modelling has been in 
finding a physical explanation for the decrease. A possible explanation 
for this effect which has sometimes been given is electron scattering in the
absorbing region, which would reduce the X-ray flux in an energy-independent
way. In the case of XBT\thinspace 0748-676, Parmar et al. (1986) suggested
that the unabsorbed component was the result of rapid chages in column density
which can mimic a soft excess. A further problem in absorbed plus
unabsorbed modelling is in explaining the variation of the normalization of
the absorbed component during dip intervals, which although approximately 
constant, always exhibits a wave (in the plot of normalization {\it versus} 
intensity) that is difficult to explain.

\medskip \noindent
The dipping LMXB sources do not, in general, exhibit the spectral evolution
in dipping expected for absorption of a single-component spectrum by cool
material, ie a hardening of the spectrum, and individual sources can display 
hardening, energy-independence or even softening, as in X\thinspace 1624-490
(Church \& Ba\l uci\'nska-Church, 1995). It has been possible to explain the
complexity of this behaviour using the two-component model
proposed by Church \& Ba\l uci\'nska-Church (1995), in which X-ray emission
originates as point-source blackbody emission on the surface of the neutron star,
or in the boundary layer at the surface, plus extended Comptonized emission,
probably from the accretion disk corona (ADC). Using this model, good fits were obtained 
to the {\it Exosat} data in the case of X\thinspace 1755-338,
for example, in which dipping appears to be mostly due to absorption of the point-source
blackbody component (Church \& Ba\l uci\'nska-Church 1993), and rapid variability
in dipping is due to the point-like nature of this component.  
However in this source, dipping is less than 20\% deep in the 1--10~keV band, 
and the implication is that the projected size of the absorbing region is 
smaller than the projected size of the Comptonizing region. 

\src\ was observed using {\it ASCA} on 1993, May 2nd. Dipping was very deep, 
reaching 100\% during every dip, showing that all
emission regions became covered by absorber, and there was clear evidence
for an unabsorbed part of the spectrum during dipping (Church et al. 1997).
The dip spectra could be well fitted by the two-component model by allowing
progressive covering of the extended Comptonized emission, and the model 
flux I can be expressed in the form:

\[
\rm {I}= \rm {e^{-\sigma N_H} \;\lbrace I_{BB}  e^{-\sigma N_H^{BB}}
+\;  I_{PL}\,(f\, e^{-\sigma N_H^{PL}}+\; (1 - f))\rbrace }
\;\;\;\;\;\;\;(1)
\]

\noindent
where $\rm {I_{BB}}$ and $\rm {I_{PL}}$ are the energy-dependent fluxes
of the
blackbody and cut-off power law components,
$\rm {N_H^{BB}}$ and $\rm {N_H^{PL}}$ are the column densities of each
component during dipping, additional to the non-dip column density $\rm {N_H}$,
and $\sigma$ is the photoelectric absorption crosss section of
Morrison and McCammon (1983).
A power law is used to approximate Comptonization at energies well
below the break energy. This fitting showed that
the point-like blackbody is absorbed as soon as the absorber overlaps the
neutron star, but the extended component is covered progressively as
dipping develops as shown by the smooth increase of the covering fraction 
{\it f} $\;$ from zero to $\sim$1.0. This is consistent with a dense absorber 
moving across an extended emission region, and differs from partial
covering of the source regions by a ``blobby'' absorber having density variations 
such that different parts of the absorber are more or less absorbing to X-rays. 
Although progressive covering and partal covering by a blobby absorber are very
different, both models can be represented by the same spectral model which
takes no account of the geometry of overlap of source and absorber, but
only of the total covering fraction.

Progressive Covering also differs substantially from the ``absorbed plus unabsorbed'' approach
previously used for fitting the group of sources with an unabsorbed spectral
component in dipping. Firstly, there is no decrease in normalization of any
component and so no difficulty in explaining either the decreasing normalization
of the unabsorbed component, or the wave in the normalization of the absorbed
component as a function of intensity as dipping develops.
Secondly, there is a clear
explanation of the unabsorbed component which is simply the uncovered part
of the extended Comptonized emission. In absorbed plus unabsorbed modelling,
the origin of the unabsorbed component was not clear, but it was often 
suggested that electron scattering
was responsible, clearly a radically different
explanation. In fitting the {\it ASCA} data on \src\ (Church et al. 1997), 
dipping was explained purely
in terms of phooelectric absorption by material with Solar abundances
without requiring electron scattering in the energy band $\sim$1--10 keV,
and it was shown from the Thomson and photoelectric absorption
cross sections that absorption strongly dominates 
over electron scattering below 10 keV so that little electron scattering is,
in fact, expected. With the same model, it has recently been possible to fit 
the spectral evolution in dipping in XBT\thinspace 0748-676 
during the {\it ASCA} PV phase observation (Church et al. 1998a). The fitting
required a line feature at $\sim$0.65 keV having the same progressive covering factor
as the extended Comptonized emission, and as the line probably originates in the ADC,
this provides strong evidence that the extended continuum emisson region is also the ADC.

\medskip \noindent
High quality data were also obtained on XB\thinspace 1916-053 with {\it
Rosat}, and we present here spectral analysis results in which
we test whether the approach applied with the {\it ASCA} data can
also describe dipping at lower energies in the 0.1--2.0~keV band of 
the {\it PSPC} instrument.

\section{Results}
XB\thinspace1916-053 was observed using {\it Rosat} for 49~hr 
starting on 1992, Oct 17,
corresponding to about 60 orbital cycles. Data were extracted by
selection in a 2~\arcmin radius circle centered on the source from
the sky image. Background data were selected from an annular region 
of the image excluding all apparent point sources. The quality of 
background subtraction was tested by performing spectral fitting with 
this background data, and without subtraction, and it was found that 
there was little effect on spectral fitting results
even if no subtraction was made. 

\begin{figure}
\epsfxsize 82 mm
\leavevmode\epsffile{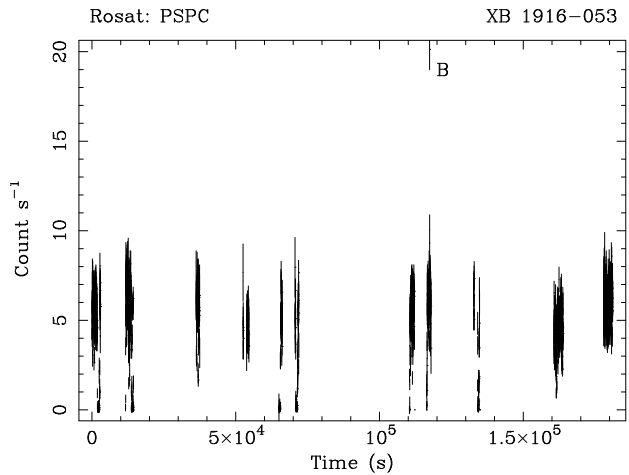}
\caption{Light curve for the total observation in the band 
0.1--2.0~keV in 16~s bins. The burst is labelled B.\label{fig1}}
\end{figure}

\subsection{Light Curves and Folding Analysis}
The background-subtracted light curve of the
observation is shown in Fig.~1 in the energy band 0.1--2.0~keV,
in which one burst is evident. The data are very fragmented, and coverage of 
dipping is not very good, however, in 3 orbits, there is complete 
coverage of the dips. Period searching was carried out using a standard
program which folds the data on trial periods, and provides the most likely
period from the turning-point of a $\chi^2$ {\it versus} period plot.
In principle, the result may be further refined by producing a model
template for one X-ray cycle which can be fitted sequentially to each cycle of the
data to test whether a time difference accumulates between data
and model. This method also provides error estimates for the X-ray period.
In the present observation in which the coverage of dipping 
is not good, this is not justified, and we quote an approximate
period uncertainty 
from the uncertainty in the peak of the $\chi^2$ - period plot due
to the errors in individual $\chi^2$ values.
The best fit X-ray period found by period searching of the 
complete observation in the band 0.1--2.0~keV was 3004$\pm 12$~s.
The period uncertainty may actually be larger than quoted as all
methods of error determination assume that the shape of dips does
not change.
The value obtained is consistent with the previous determinations:
3015$\pm $17~s (Smale et al. 1986), 3005$\pm$6.6~s (Smale et al. 1992)
and 3005$\pm$20~s (Church et al. 1997).
To show the folded light curve, only the 3 orbits with good coverage 
of dipping were included, since
if all of the data is included, dipping in the folded light curve is 
not 100\% deep due to averaging of dips that have some differences in phase
and shape, and does not reflect dipping in a typical dip.

Folded light curves in
the two energy bands 0.1--1.0~keV and 1.0--2.0~keV are shown in 
Fig.~2. It can be seen that dipping is essentially 100\% deep 
in both bands, and that the intensity remains close to zero 
during almost all of the dip until dip egress begins.
This contrasts with dipping in the 1--10~keV band as revealed by {\it
ASCA}, in which the intensity falls rapidly to $\sim $ 10\% of the
non-dip level, but then more slowly, so that the intensity touches zero
for only a short time at the deepest part of the dip (Church et al. 1997).
This suggests that only at the
deepest point in each dip is there complete overlap between the high
density central regions of the absorber and the source.
In the {\it Rosat PSPC} band, a smaller column density will
lead to zero intensity and so there will be an extended period 
of deep dipping. Moreover, the lack of variation between the depth of 
dipping in the two {\it Rosat} energy bands is consistent with a
high column density in the absorber, as was the case during the {\it ASCA}
observation. When the hardness
ratio formed from the two folded light curves is examined, there is
little change in hardness ratio from the non-dip value at dip
ingress or egress. 
In {\it ASCA}, there was an increase in hardness defined
in terms of the bands 2.0--12.0~keV and 0.5--2.0~keV during dip ingress
and egress, as dipping was not 100\% in the dip transitions, and not equal
in these bands (Church et al. 1997). Finally, it can be seen that there
appears to be asymmetry between dip ingress and egress; however this must be
treated with caution because of the limited amount of dip data
included in the folding.

\begin{figure}
\epsfxsize 82 mm
\leavevmode\epsffile{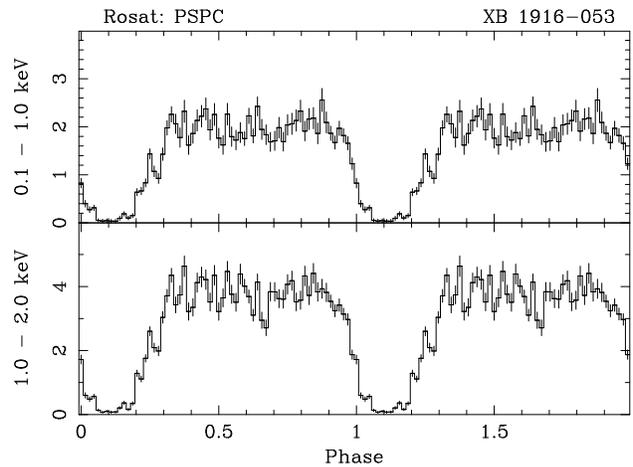}
\caption{Light curves in the bands 0.1--1.0~keV and 1.0--2.0~keV folded
on the period of 3004~s.\label{fig2}}
\end{figure}

\subsection{Spectral Evolution in Dipping}
The good quality of the {\it Rosat} data has allowed detailed spectral
analysis to be carried out for non-dip data and for dip data selected
in several intensity bands. Non-dip data was selected 
from the complete observation, but excluding the burst,
from the 
intensity band 5.5--7.0~count s$^{-1}$, and 4 intensity bands: 4.0--5.5, 
2.5--4.0, 1.0--2.5 and 0.0--1.0 count s$^{-1}$ were used
for dipping. Background was subtracted, systematic errors of 2\%
were addded and the spectra regrouped appropriately, to a minimum
of 100 counts per channel for the non-dip spectrum, and to minima of 80,
60, 40 and 40 counts for the 4 dip spectra (with decreasing intensity). 
The solar abundances of Anders and Grevesse (1989) were used throughout.
Firstly, simple one-component spectral models were tried
such as an absorbed power law, absorbed bremsstrahlung and absorbed
blackbody models. In all model fitting, the emission parameters were fixed
in fitting dip spectra at the non-dip values, since dipping is due to
absorption, not source changes; thus for the simple blackbody model
the blackbody temperature and normalisation were fixed.
This model gave a marginally unacceptable 
fit to the non-dip spectrum, whereas the bremmstrahlung model
gave an acceptable fit. However, both models were quite unable 
to fit dip data, giving reduced $\chi^2$ values of 3.98 and 3.25 
respectively for a simultaneous fit to non-dip and 4
dip levels. The power law model gave a good fit to the non-dip data,
but was also unable to fit dip spectra because of strong excesses 
at low energies of the 
model over the data. Thus dipping cannot be modelled by simple absorbed
models, and it was clear that an `unabsorbed component' was present, 
ie part of the non-dip emission persisted at low energies in dipping and
was not absorbed. 
We next tried the two-component model with progressive
covering of the extended emission which gave a good fit to the {\it ASCA}
data, represented by equation 1. The point-like blackbody is assumed to
be covered immediately the absorber overlaps the emission region, ie the
neutron star.

The {\it ASCA} results (Church et al. 1997) showed that the blackbody
component had $\rm {kT_{BB}}$ = 2.1~keV, peaking outside the {\it Rosat}
band and so the flux of this component in the 0.1-2.0~keV band is
expected to be small, assuming the source did not change markedly between 
the observations, making it difficult to determine blackbody parameters.
Consequently, we first tried the above model omitting the blackbody, and
then repeated the fitting with the blackbody added. The results for the
power law component did not change markedly, and sensible values for 
blackbody parameters were obtained. In this case, it is clearly better
to present results from the fitting with the blackbody present than
with it omitted, or with blackbody parameters set to the {\it ASCA} values.

Firstly, simultaneous fitting was carried out to the non-dip spectrum 
and all dip levels, and good fits were obtained.
Best fit values of $\rm {kT_{BB}}$ = 1.95$\rm {^{+0.74}_{-0.34}}$~keV, 
and power law photon index $\Gamma $ = 2.32 $\rm {\pm
0.04}$ were obtained, and the column density for the non-dip spectrum = 
$\rm {3.9\pm 0.1\times 10^{21}}$ H atom $\rm {cm^{-2}}$. 
Very similar values were obtained by fitting the non-dip spectrum alone. 
Although $\Gamma $ appears to be well-determined, the band of the {\it PSPC}
is relatively narrow for determining power law index values (see Sect. 3).

Parameters for the 4 dip spectra were
further optimized by fitting these individually keeping $\rm {kT_{BB}}$,
$\Gamma $ and the normalizations at the best-fit values. 
The results are shown in Fig.~3 and Table~1, 
which gives the blackbody and power law
column densities $\rm {N_H^{BB}}$ and $\rm {N_H^{PL}}$ during dipping 
{\it additional} to the non-dip column density, and the progressive
covering fraction {\it f}. The column densities and the covering fraction
{\t f} are plotted a function of X-ray intensity in the band 0.1--2.0~keV
at different stages of dipping in Fig.~4.
These
results show that during dipping, the power law emission is progressively
covered until in the deepest dip spectrum, {\it f} is 97\%
and the power law column density has risen to $>$$\rm {10^{23}}$ 
H atom $\rm {cm^{-2}}$. However, the covered part of the power law 
emission is quickly removed to energies greater than 2~keV, 
and we have the interesting situation that
the {\it PSPC} effectively only detects and measures the properties of
the uncovered, ie the unabsorbed part of the spectrum. The
blackbody suffers relatively little absorption until the partial covering
fraction is $\sim $ 50\%, at which point it increases rapidly, and by the
second level of dipping (in the band 2.5--4.0 count s$^{-1}$, the 
blackbody flux in the {\it PSPC} is close to zero, and for 
the remaining levels of dipping, it is zero.
The larger values of $\rm {N_H}$ for the
blackbody are expected since this point source emission
effectively monitors regions towards the centre of the absorber,
whereas the extended Comptonized emission 
gives an $\rm {N_H}$ integrated across
the absorber. The lack of change in column density for the blackbody
until {\it f} becomes appreciable might also be expected in terms
of our physical picture of an extended absorber which would have to
cover part of the ADC before reaching the neutron star.

\begin{figure}
\epsfxsize 82 mm
\leavevmode\epsffile{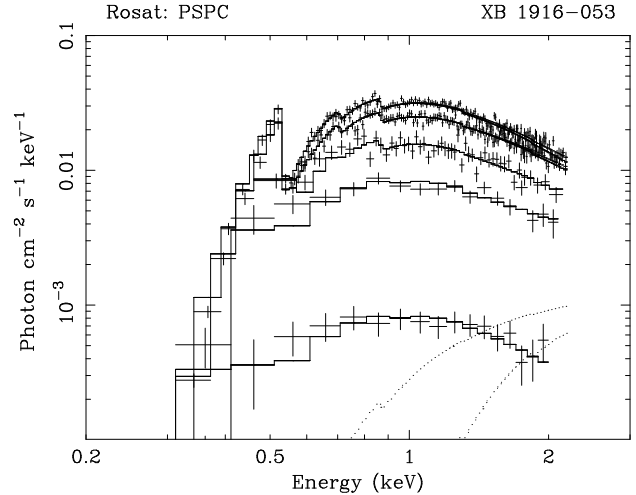}
\caption{Spectral fitting results for non-dip and 4 levels of dipping.
The total model is shown as a stepped line, and the blackbody
contribution (only visible for the non-dip and first level of dipping)
is shown dotted.\label{fig3}}
\end{figure}

Although the {\it PSPC} is not ideally
suited to determining values of $\Gamma $, it is ideal for investigating
the spectrum in the region of the low energy cut-off, and comparison of 
the non-dip and dip spectra shown in Fig.~3 reveals that there is no 
detectable change in the low energy cut-off, which is strong evidence 
that the unabsorbed part of the spectrum is the uncovered part of the 
non-dip emission.

\begin{table}
\caption{Best fit spectral fitting results. Intensities {\it I} are in
count s$^{-1}$, and $\rm {N_H}$ values are column densities additional
to non-dip values and are in units of $\rm {10^{22}}$ H atom $\rm {cm^{-2}}$.}
\halign{#\hfil&&\quad#\hfil\cr
&$\rm {I}$ & $\rm {N_H^{BB}}$ &$\rm {N_H^{PL}}$ & f & $\rm {\chi^2/dof}$\cr
\noalign{\medskip\hrule\medskip}
&5.5 - 7.0  & 0.0             & 0.0        & 0.0            & 154/165\cr
&4.0 - 5.5  & 1.4$\rm {^{+2.9}_{-0.9}}$    & 2.5$\rm {^{+1.0}_{-0.5}}$ & 0.208$\pm $0.020 & 146/153\cr
&2.5 - 4.0  & $>$20           & 4.2$\pm $1.8 & 0.503$\pm $0.014  &53/47\cr
&1.0 - 2.5  & $>$20           & 5.0$\pm $1.6 & 0.736$\pm $0.013  &24/18\cr
&0.0 - 1.0  & $>$20           & $>$20        & 0.974$\pm $0.003  &13/18\cr
\noalign{\medskip\hrule\medskip}}

\end{table}

\begin{figure}
\epsfxsize 82 mm
\leavevmode\epsffile{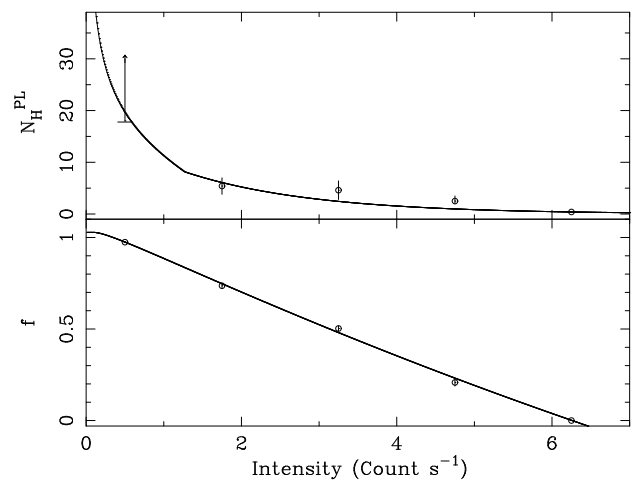}
\caption{The covering fraction {\it f} and column density
$\rm {N_H^{PL}}$ in units of $\rm {10^{22}}$ H atom cm$^{-2}$
of the extended power law emission component as a function of source
intensity. The solid lines show polynomial fits to the data points.
\label{fig4}}
\end{figure}

\section{Discussion}
The LMXB source \src\ was observed with {\it Rosat}, and during this
observation dipping was 100\% deep at all energies in the {\it PSPC}
band, showing that all emission regions were completely covered by
the absorbing region. The {\it PSPC} spectra are well fitted 
by the two-component model proposed by Church and Ba\l uci\'nska-Church 
(1995) consisting of point-like blackbody emission from the surface or
boundary layer at the surface of the neutron star, plus extended
Comptonized emission from the ADC. Good fits to the dip spectra are
obtained by allowing progressive covering of the extended emisson
component, with the covering fraction rising smoothly from zero to 
close to unity. This, together with the large vaues of column density
determined for the dip spectra, demonstrates that the absorber is large
and dense, and moves progressively across the extended source region.
Thus the covering is not partial in the sense of a ``blobby'' absorber,
ie an absorber partially transmitting to X-rays
between high density regions. The blackbody emission, on the other hand,
is very rapidly absorbed once the absorber covers the neutron star.
The high quality {\it PSPC} spectra show 
clearly that there is no detectable change in the low energy cut-off of 
the spectrum from non-dip to any of the stages of dipping, strongly 
supporting the contention that the unabsorbed emission is simply the 
uncovered part of the extended ADC emission. This approach differs 
radically from the ``absorbed + unabsorbed'' modelling, often used 
previously for this source and similar sources in which the nature
of the unabsorbed peak, or soft excess, was seen as being due to
electron scattering or possibly fast changes of column density 
mimicing a soft excess. In the progressive covering model, there
is a simple explanation of the unabsorbed component as uncovered emission.
The progressive covering model also differs
in its application, since the normalizations of the spectral components
are fixed, and the difficulty in explaining the changing normalizations
in absorbed plus unabsorbed modelling does not exist.
With the progressive covering model, it is not necessary to invoke
an explanation for the unabsorbed component such as being due to electron
scattering, and dipping is explained entirely in terms of
phtoelectric absorption. We do not need to invoke electron scattering
in fitting either the {\it Rosat} or the {\it ASCA} data, and calculations
of the relative loss of X-ray intensity by the two processes: electron
scattering and photoelectric absorption show that absorption strongly
dominates at most energies below 10 keV. However above 10 keV,
the Thomson cross section becomes larger than that for
photoelectric absorption and dipping above 10 keV may
involve a degree of electron scattering.

From the results shown in Table~1, the covering fractions {\it f} 
may be used to calculate the relative energy fluxes of the
absorbed and unabsorbed parts of the spectrum, ie the covered and
uncovered components. In the band 0.1--2.0~keV,
the percentage of the absorbed part was found to have the values:
0\%, 2.6\%, 3.2\%, 5.6\% and 0.1\% for the sequence of 5 spectra
going from non-dip to deep dipping. As the source progresses into dipping,
the effective normalization of the absorbed power law component ({\it f}
times the actual normalization) increases as {\it f} increases.
$\rm {N_H}$ also increases, and so the percentage of this component peaks
at about 6\%. Consequently, the {\it PSPC} acts as a very good filter in
allowing measurement of the unabsorbed part, essentially alone. The main
consequence of this is that the most important parameter determining the
{\it PSPC} spectra is {\it f}; column densities do not affect the 
uncovered part of the power law emission, and hence {\it f} can be
determined with good accuracy. Thus the spectral evolution in dipping is
determined in the {\it Rosat} band
by the geometric overlap between the extended source component
and the absorber, and the values of {\it f} can be used in principle to
determine the shape of the absorber for given assumed simple shapes of
the ADC. To do this, values for {\it f} should be obtained 
as a function of orbital phase, not intensity. However,
allowing for the need for a minimum of 7 phase bins to cover non-dip,
ingress, deep dip and egress,
the quality of the present data do not allow us, for example,
to discriminate between geometric models for the ADC.

The power law index was found to have a value of 2.3, 
however it should be borne in mind that the band of energies
available in the {\it PSPC} from which an index can be obtained is
very narrow. Recent work on the source using 
the very broad band of {\it BeppoSAX}
of 0.1--300~keV (Church et al. 1998b) has shown that the non-dip spectrum
is well fitted by a blackbody plus cut-off power law model, with
a cut-off energy of $\sim$80~keV. The underlying power law photon index
may, of course, be well determined using this very wide band, and it
was shown that with more restricted energy bands, there is a tendency
to overestimate values of the index. Using the LECS and MECS bands
without the other instruments, ie in the energy range
0.1--10~keV, $\Gamma $ $\sim $1.75 whereas in the total band extending
to $>$100 keV, $\Gamma $ $\sim $1.6. A similar effect, or larger,
may be expected in the present work. However, 
it is clear that the progressive covering model gives
a good explanation of spectral evolution in dipping in XB\thinspace
1916-053, and has also been shown to describe dipping well in
XBT\thinspace 0748-676 (Church et al. 1998a), and it represents 
an improvement in our understanding of these sources. An interesting
question that remains is why do some sources such as XB\thinspace
1916-053 and XBT\thinspace 0748-676 have the extended emission component
often totally absorbed, whereas in others, for example, X\thinspace
1755-338, there is apparently little absorption of this emission.

\end{document}